# Disordered Cellulose-based Nanostructures for Enhanced Light-scattering

Soraya Caixeiro[a*], Matilda Peruzzo[a,¥], Olimpia D. Onelli[b], Silvia Vignolini[b*] and Riccardo Sapienza[a]

*a Department of Physics King's College London, Strand London WC2R 2LS, UK.*
*b Department of Chemistry, Cambridge University Lensfield Road, Cambridge CB2 1EW, UK.*
*¥Current address: Institute of Science and Technology Austria (IST Austria), 3400 Klosterneuburg, Austria*
*\*Corresponding authors: soraya.carlos_caixeiro@kcl.ac.uk, sv319@cam.ac.uk.*

Cellulose is the most abundant bio-polymer on earth. Cellulose fibres, such as the one extracted form cotton or woodpulp, have been used by humankind for hundreds of years to make textiles and paper. Here we show how, by engineering light matter-interaction, we can optimise light scattering using exclusively cellulose nanocrystals. The produced material is sustainable, biocompatible and, when compared to ordinary microfibre-based paper, it shows enhanced scattering strength (x4) yielding a transport mean free path as low as 3.5 µm in the visible light range. The experimental results are in a good agreement with the theoretical predictions obtained with a diffusive model for light propagation.

With the term "paper" we include a large variety of cellulose-based composite materials that find use in everyday life such as packaging and printing. Recently, paper-based technologies have captivated increasing interest not only due to their applications in sensing[1-3] and lasing,[4] but also in 3D-cell scaffolding.[5] Cellulose can be easily functionalised to produce materials with enhanced mechanical, optical, and chemical properties due to its intrinsic fibrillary morphology and consequent porosity. These new materials are particularly attractive from an industrial point of view thanks to their low production costs.[6]

The main component of paper is cellulose.[7] Natural cellulose can be extracted from different sources: ranging from plants (such as wood pulp or cotton) to bacteria, to invertebrates and some marine animals,[8] nonetheless it is consistently found to have fibrillary nature.[9] Such natural fibres are generally composed of amorphous and crystalline regions, see figure 1(a). In paper-manufacturing process, moist cellulose fibres extracted from natural sources are compressed together and dried. The thickness of the fibres and their packing density determine the optical response of the material.[7] Conventional fibres in paper are several tens of microns in diameter and therefore, they are not ideal to produce a strong scattering response, figure 1(b-d). By acid hydrolysis such crystalline region, called cellulose nanocrystals (CNCs) can be extracted and suspended in water.[8] CNCs can be considered as rod-shaped colloidal particles typically 150 nm in length and a few nm in diameter [10] (figure 1e), with a significantly high refractive index (about 1.55 in the visible range). CNCs have been received an increase interest in photonics, due to their colloidal behaviours and their ability to self-assemble into cholesteric optical films[11, 12].

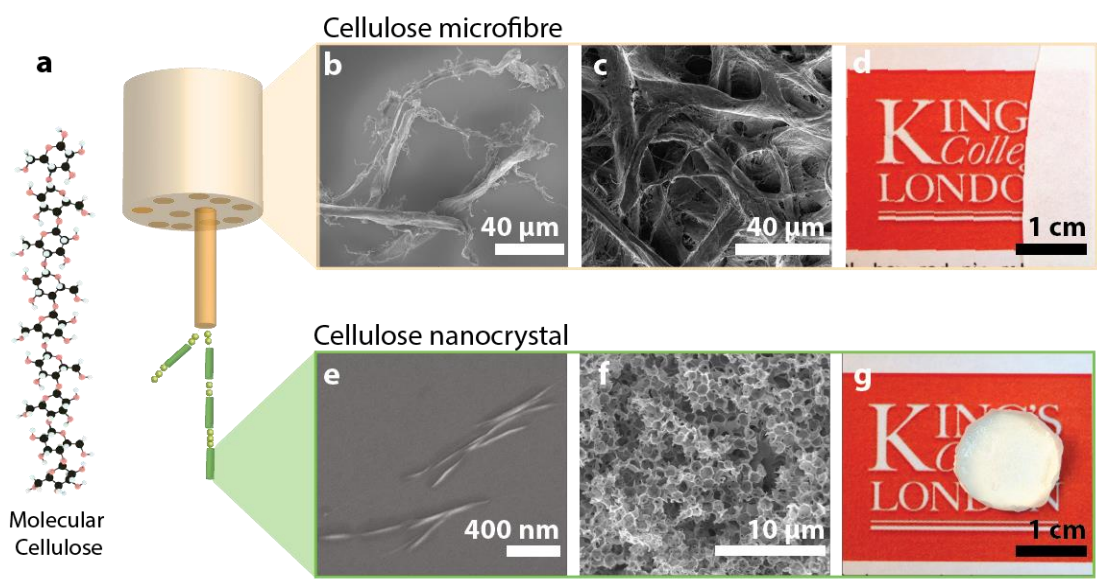

**Figure 1 Structure and fabrication-** A diagram showing the hierarchical structure of a cellulose chain is illustrated in panel a. A cellulose fibre is composed of fibrils (orange cylinder) with alternating crystalline (dark green rods) and amorphous (light green spheres) sections. On the top, cellulose fibres (panel b) used to fabricate white paper (panel d) whose fibrous structure is shown in the SEM image (panel c). On the bottom, cellulose nanocrystals (panel e) which can be self-assembled in the shape of a photonic glass structure by polystyrene sphere templating (panel g), and the respective SEM image (panel f).

However, while CNCs have been intensively studied for structural colour applications, [13] such materials have never been exploited to maximise scattering. Maximal scattering strength is, in fact, challenging to obtain: optimal scattering design comes from the balance of scatterers' size, refractive index contrast and filling fraction. Therefore, it is fulfilled for dielectric particles of diameters comparable with the wavelength of light packed with maximal density assuring also a high refractive index contrast between the scatters and their surrounding environment.

Scattering inside paper is measured via the transport mean free path ($\ell_t$), the length beyond which the propagation is random, which for paper is typically of the order of 20 μm.[14] Maximal scattering, which means minimal $\ell_t$, is an important technological goal for producing whiter and more opaque materials. More efficient scattering implies that a smaller quantity of material is needed to achieve the same white coating.

Here, we report the bottom-up fabrication technique for the production of a new scattering paper-like material. Unlike conventional paper, our starting material is the smallest constituent of cellulose: the cellulose nanocrystal[11] (figure 1a). In contrast, we produce a nanostructure made solely of CNCs capable of improved light-matter interactions, due to its much smaller feature size (figure 1e-g).

By characterising the scattering response of the CNC-based photonic glass we obtain a scattering more than 400% stronger than for standard cellulose fibre paper. The experimental results compare well with a diffusive model. Furthermore, we estimate the optimum fabrication conditions for maximum scattering and opacity, and point out a possible strategy to minimise costs.

A cellulose inverse photonic glass [15,16] is fabricated using a templating technique that consists in the co-deposition of monodisperse PS spheres and CNCs and sub-sequential chemical etching of the PS spheres. This geometry is particularly convenient to optimise light-matter interaction because provides the right balance between the size of the scattering elements (at the edge of the spherical voids), and a high filling fraction.[16] Commercial cellulose nanocrystals (Forest Product Laboratory Canada) are extracted by sulfuric acid treatment of wood-pulp, leaving negative charged sulfate half-esters which are neutralised with $Na^+$ ions. The dimension of the colloids is around 5 nm in diameter and range from 150 to 200 nm in length. The pH of the suspension is neutral, while the surface change is 278±1 mmol/kg estimated by conductometric titration.[17] The scanning electron microscope (SEM) image in figure 1e shows the characteristic needle-like geometry of the CNCs.

A 4 %wt CNCs aqueous solution (deionized water) is mixed with colloidal monodisperse polystyrene (PS) spheres of diameter d = 1.27 μm (Micro particles GmbH), such that the dry weight ratio between CNCs and PS spheres is 2:3 respectively. The obtained suspension is then cast into a hollow Teflon cylinder attached to a glass substrate with PS as in reference.[18] Prior to this, the Teflon cylinder is immerged in a NaOH bath to improve the hydrophilicity of its surfaces, while the glass is coated with PS to stabilize the film (to avoid cracking during drying).

The samples are kept in partially sealed containers and dried for 1-2 weeks in a quasi-saturated water vapour atmosphere kept at a constant temperature (30 °C). Such conditions allow a slow evaporation rate which further improved the film quality by avoiding cracking and delamination. Once the sample is dry, the PS spheres are selectively etched in a bath of toluene for approximately 3-9 hours, depending on the sample thickness (50-500 µm). Toluene also removes the PS coating from the glass substrate and separates the Teflon cylinder from the glass; this facilitates the detachment of the sample, yielding a free-standing cellulose inverse photonic glass, a nanostructured paper (figure 1g). The CNCs properties are unaffected by the toluene bath,[19,20] as confirmed by the transmission experiments conducted for a timespan of over 10 hours, showing no significant transmission change.

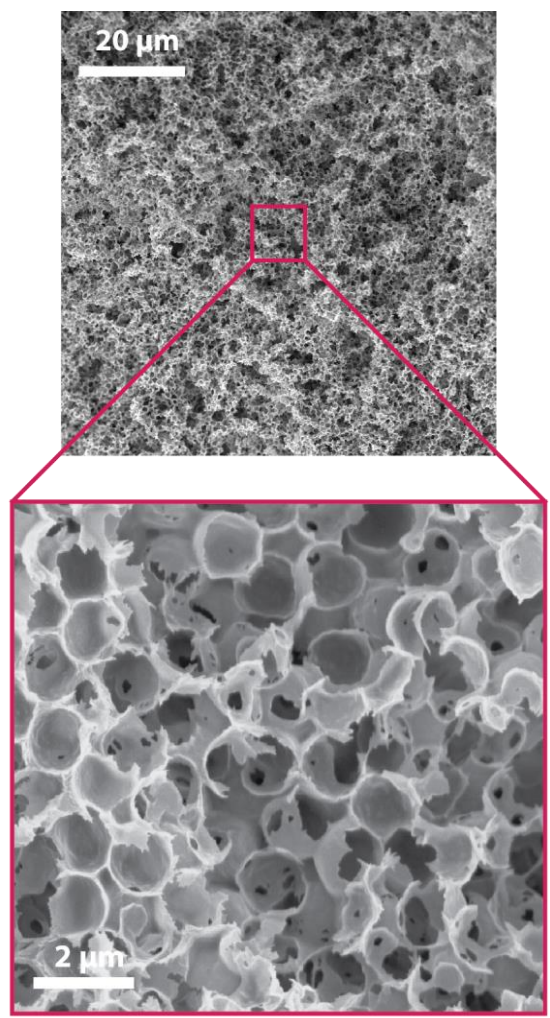

**Figure 2 Sample morphology** – The panel on the top reports an overview of the sample showing it is porous on a large scale. When imaged at higher magnification (below), it is possible to observe the micrometre-sized voids formed around the etched spheres. Smaller pores connecting the voids are visible, formed when the PS spheres were touching. The larger voids visible in the image are due to the sample cutting used for SEM imaging.

After the drying process, a random close-pack arrangement is formed, confirmed by SEM inspection (figure 1f) and optically by the lack of iridescence and enhanced normal reflection. The only observable change is a minor reflection from the surface in contact with the glass i.e. where the cellulose layer is more compact. The resulting cellulose inverse photonic glass is shown in figure 2. An SEM image of the structure reveals spherical voids of about 1.3 µm (where the PS spheres were present prior to etching), as well as circular openings characteristic of a close packed structure, (in correspondence of the position where the PS spheres touched each other before template removal). We observe that such topology is homogenous throughout the sample, as further confirmed by transmission studies on different areas of the sample (see later). In addition, a photograph of one of the samples fabricated is shown in figure 1g (approximately 1.5 cm in diameter and 100 µm thick): the increased opaqueness of the photonic glass paper is visible even by the naked eye, when compared to conventional paper of similar thickness (figure 1d).

We compare the scattering properties of CNCs photonic glass and common cellulose fibre paper by measuring $\ell_t$, by means of total transmission measurements (T) performed with an integrating sphere which collects the transmitted flux over all angles.[20] The measured light is sent to a spectrometer which provides spectral information. The photonic Ohm's law,[21,22] which is described by the change in total transmission (T) as a function of the sample thickness (L), is obtained via the stationary solution of the diffusion equation (assuming a slab geometry)[15, 21]:

$$T(L, \lambda) = \frac{1}{\alpha z_e} \frac{\sinh[\alpha(z_p+z_e)] \sinh(\alpha z_e)}{\sinh[\alpha(L+2z_e)]}, \quad (1)$$

where $\alpha = 1/\ell_a$ is the reciprocal of the absorption length $\ell_a$, $z_e$ is the extrapolation length and $z_p$ is the penetration length, typically taken to be equal with $z_e = \frac{1}{2\alpha} \ln\left(\frac{1+\alpha z_0}{1-\alpha z_0}\right)$ and $z_0 = \frac{2}{3} \ell_t \frac{1+R}{1-R}$. R is the averaged reflectivity (R = 0.39 assuming a filling fraction of ≈55% and n= 1.55).

The microfibre paper used is Whatman® filter paper, Grade 1 with a reported thickness of 180 µm confirmed by SEM inspection. While for the paper photonic glass it was possible to produce samples of different thicknesses, for cellulose fibre paper multiple sheets of paper had to be compressed together in order to increase the overall thickness. The transmission spectra of cellulose fibre paper, averaged over 3 measurements, were fitted with equation (1) for each wavelength as shown in Figure 3a and b. A typical fit at λ = 600 nm is plotted in figure 3a highlighting the exponential dependence of the inverse transmission (1/T) on the thickness (L) due to absorption. Here, $\ell_t$ and $\ell_a$ were estimated by a two-parameters fit of equation (1). We have used a multi-step fitting routing: 1. $\ell_t$ and $\ell_a$ were taken as free parameters of the fit to obtain approximate values for each wavelength; 2. each parameter was fitted independently using the other parameter as an input, until convergence was achieved (after 4 iterations). Furthermore, the fitted value from each wavelength was used as the starting point of the consequent one to improve the convergence. The measured $\ell_t$ ranges between 15 and 22 μm in the visible spectra as depicted in figure 3b), while $\ell_a$ is of the order of a millimetre. The measured $\ell_t$ is an underestimation of the actual value, although the air gaps between the sheets are expected to increase the apparent $\ell_t$, we estimate by microscopy inspection that they are less than 10% of the sample thickness. The reflection at each interface (about 40% for each sheets), on the other hand, is a more significant effect which increases the total reflectivity thus lowering the transmission and increasing the measured $\ell_t$.

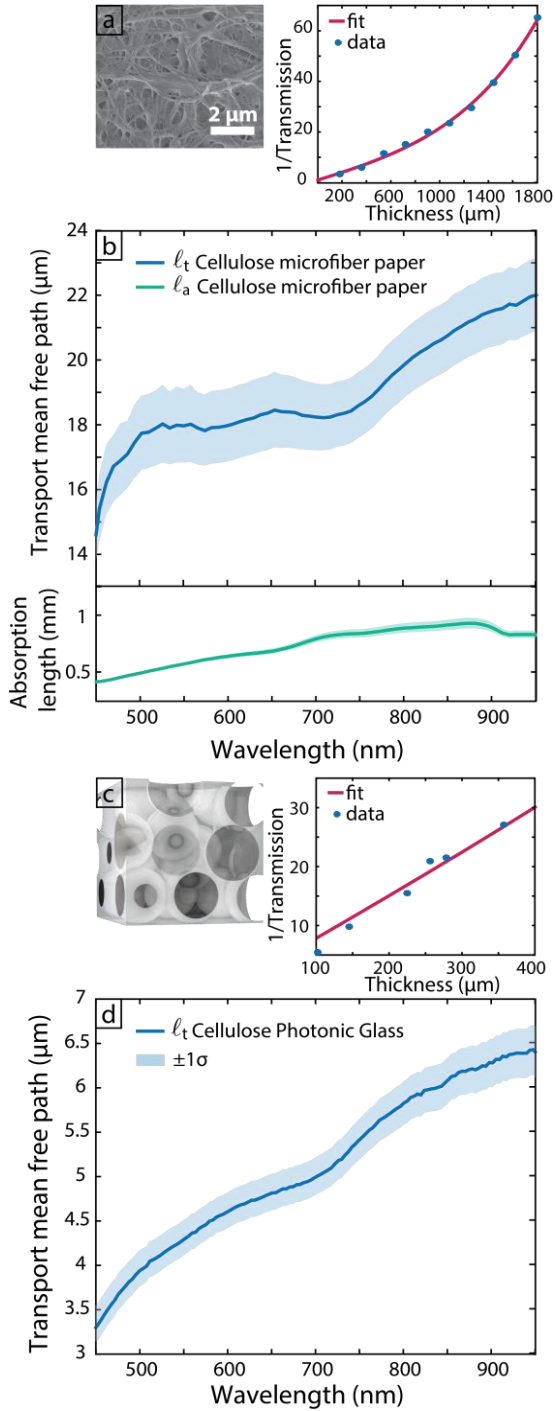

**Figure 3 Measured scattering strength** - a) Cellulose microfibre paper SEM and fit of the total transmission data at different thicknesses (λ = 600 nm). b) Measurement of the light transport mean free path (blue) and absorption length (green) for microfibre paper. The error bars are shown by the shaded area around the lines, for both fitting parameters. c) Model of structure of a cellulose inverse photonic glass and fit of the total transmission data at different thicknesses (λ=600 nm). d) Measurements of the light transport mean free path for the inverse cellulose photonic glass. The error bars are shown by the blue shaded area around the line.

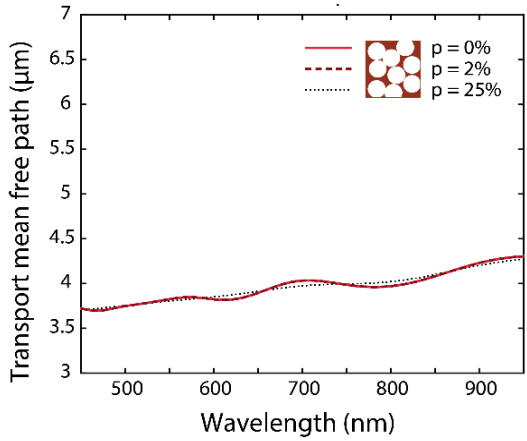

**Figure 4 Modelling experimental results -** Theoretical transport mean free path calculated for air spheres in cellulose matrix $n_{cellulose}$ = 1.55 at a filling fraction of f = 0.55 as a function of wavelength (full red line) assuming polydispersity of 2% (dashed red line) and assuming a polydispersity of 25% (dashed black line).

The measurement of $\ell_t$ in the case of the photonic glass paper was performed by comparing samples with different thicknesses (figure 3c) in the range 100 – 400 μm. Such thicknesses are estimated by SEM. The values of the thickness were averaged over different areas on the sample, with an error of around 5%. The transmission spectra of the cellulose photonic glass were averaged over different regions of the sample and a dispersion less than 5% is measured. Using the same procedure depicted above, the data was fitted with Equation (1). As $\ell_a$ in these samples is much larger than the sample thickness, lossless Ohm's law is valid, as shown in figure 3c. Therefore, for simplicity and stability of the fit, we have chosen α=0. Figure 3d shows $\ell_t$ obtained as a function of wavelength. The statistical error of the fit accredited to minor sample-to-sample variation is estimated to be less than 10%. As expected, $\ell_t$ decreases towards shorter wavelengths as expected by Mie theory (see later). The measured $\ell_t$ is in the range $\ell_t \approx$ 3-7 μm for the visible range with very shallow resonances. The lack of resonances is expected, as air voids in a higher refractive index matrix are poor resonators, in contrast with high refractive index spheres which show appreciably stronger resonances[15,18]. The scattering strength of the photonic glass paper is significantly stronger: $\ell_t$ is 4 times smaller than that measured for cellulose fibre paper.

The theoretical calculations are performed via Mie theory and independent scattering approximation, taking into account the polydispersity of the PS spheres.[23,24] Comparing to previous works [16] we expect the photonic glass paper to have a filling fraction around f = 50-55%, smaller than the theoretical limit for hard-sphere random packing.[25] The comparison between the theoretical results for different degrees of polydispersity is shown in figure 4. Since the resonances are weak, they are unaffected by the small polydispersity.

We use the Mie model to investigate the optimum void diameter required to maximise scattering. Figure 5 plots $\ell_t$ for different void diameters (at a wavelength of λ = 600 nm), both in the absence of polydispersity and at 2% polydispersity. At d = 250 nm, $\ell_t$ is at its lowest value, around 1.27 μm while for smaller diameters $\ell_t$ increases rapidly for decreasing *d*, as dictated by Rayleigh scattering. Although scattering may be increased by using smaller PS sphere to nanostructure the CNCs, smaller CNCs than the ones used here are required to ensure maximal close-packing of the sphere and consequently stronger scattering and a lower $\ell_t$.

Polydisperse PS spheres are cheaper and easier to produce than its monodisperse counterpart, therefore we explore here the effect of polydispersity in the templating matrix. Our calculations show that even for large polydispersity, as high as 25%, the average value of $\ell_t$ is unaffected, only the resonances are damped, as shown by the dotted black line in figure 4 and 5.

In conclusion, we have presented a highly scattering nanostructured CNCs paper with $\ell_t \sim$ 3 - 7 μm.

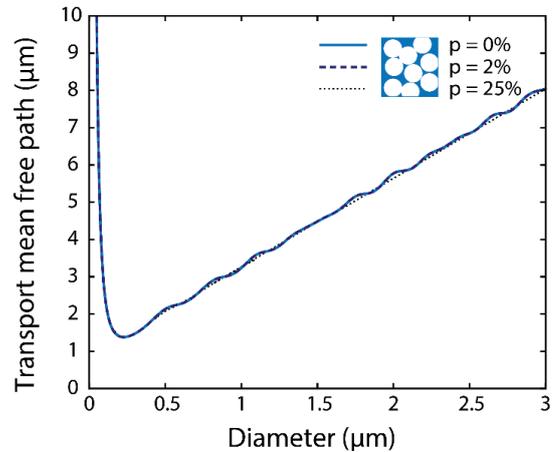

**Figure 5 Theoretical scattering strength with different void diameter -** Theoretical transport mean free path calculated for air spheres in cellulose matrix $n_{cellulose}$ = 1.55 at a filling fraction of f = 0.55, as a function of different void diameters at λ = 600 nm.

The inverse photonic glass made of CNC scatters 4 times more than standard cellulose fibre paper. By post-treatment of the film, or by adding other materials in suspension, the properties of the produced photonic glass can be further improved in terms of mechanical properties and moisture-resistance.[26-29] Increased scattering implies that the same visual contrast and whiteness can be achieved in a thinner sample. With a simple theoretical model, we identify the optimum sphere diameter of about half the light wavelength, for which the scattering strength can be maximised. Large scattering strength allows for larger contrast in sensors, thinner paper, which would reduce coating and packaging and larger porosity. Furthermore, nano-photonic enhanced paper offers the additional benefit of large porosity together with increased light-matter interaction.

## ACKNOWLEDGMENT

The authors wish to thank Michele Gaio, Giulia Guidetti and Bruno Frka-Petesič for fruitful discussions. This research was funded by the EPSRC (EP/M027961/1), the Leverhulme Trust (RPG-2014-238), Royal Society (RG140457), the BBSRC David Phillips fellowship (BB/K014617/1), and the European Research Council (ERC-2014-STG H2020 639088). The data is publicly available in Figshare.[30]